\documentclass[a4paper,onecolumn]{agn6}
\usepackage{natbib}
\usepackage{graphicx}
\usepackage[a4paper]{hyperref}
\idline{}{1}
\begin{document}
   \title{The environment of Seyferts, Liners and HII galaxies in a complete AGN sample
}
   \author{V.Zitelli \inst{1},  B.Kelm \inst{2}, P.Focardi \inst{2}
          \and
          S.Montanaro \inst{2} 
}
   \institute{INAF-Osservatorio Astronomico di Bologna,
via Ranzani  1, 40127 Bologna 
%\email{valentina.zitelli@bo.astro.it} 
              \and Dip. di Astronomia, Universit\`a di Bologna,
via Ranzani 1, 40127 Bologna, Italy
             }
\abstract{
We use a complete AGN sample (Ho et al. 1997) to study the environment 
of Seyferts, LINERs and HII galaxies. For each AGN we search for companions 
in the UZC redshift catalogue and compute local as well as large scale 
neighbour density and distance to the nearest neighbour. 
We find that on small scale ($\sim$ 0.2 $h^{-1}$ Mpc) 
LINERs exhibit denser environments than Seyferts and HII galaxies 
at 3$\sigma$ significance level. The same result is not confirmed  
when densities are computed on large scales.  
LINERs also exhibit closer nearest neighbours than Seyferts and  HII 
galaxies (at 2$\sigma$ c.l.). 
However, when excluding AGNs in early type galaxy hosts, 
the neighbour density characteristics of LINERs Seyferts and HII galaxies 
turn out to be similar, a result that confirms that the 
excess of neighbours around LINERs is most likely due to a morphology-density effect. }
   \authorrunning{V.Zitelli et al.}
   \titlerunning{AGNs and environment}
   \maketitle
%
%________________________________________________________________
\section{Introduction}
It is not yet fully understood whether the dynamical effect exerted by 
interaction with other galaxies is one of the major cause of the efficient 
gas fueling observed in AGNs. Indeed, the claimed excess of companions 
in Seyfert galaxies (Dahari 1984, Laurikainen \& Salo 1995) is not confirmed 
 \citep{Bushouse87,DeRob98} at better than 95\% significance level. 

Examination of the frequency of the occurrence of Seyferts in pairs and 
groups (Kelm et al. 1998)  has further shown that Seyferts constitute 
a $\sim$2\% in all galaxy systems, whereas the fraction of Seyferts in 
UZC-Compact Groups (CGs) turns out to be 3 times as large as among isolated 
UZC galaxies (Kelm et al. 2004). 
Interestingly, no significant differences in dynamical properties are found, 
when comparying galaxy systems which host, or not, a Seyfert member. 
For Seyferts, it has also been shown (Malkan et al. 1998) that there is 
little direct evidence for unusually high rates of interaction; less than 
10\% show tidal feature or multiple nuclei. 
And Seyferts in UZC-CGs do not seem to be more likely than other CG galaxies 
to display major interaction patterns (or a bar), which  is consistent with 
the results of earlier papers (Keel 1996, Ho et al . 1997) all finding no 
clear relationship between the presence of AGNs and detailed morphological 
properties.  

Concerning environmental differences between LINERs and Seyferts 
Kauffmann et al. (2004) have used a complete sample of galaxies 
drawn from the SDSS to suggest that the fraction  of low luminosity AGN 
(LINERs) depends very little on local density, whereas the fraction of strong 
AGN in massive galaxies decreases as a function of density. 
And Schmitt (2001) has shown that LINERs exhibit a higher percentage of 
companions than Seyferts and HII galaxies, but that differences disappear when 
considering only galaxies of similar morphological type.  
\section {The sample}
The AGN sample we use here is selected from the master list of 503 nearby galactic nuclei by Ho et al. (1997), obtained from the revised version of the 
Shapley-Ames Catalog of Bright Galaxies (RSA) and Second Reference Catalog of 
Bright galaxies (RC2). It is a flux limited catalog of galaxies with magnitude 
$B_T$ $\leq$ 12.5.
Our AGN subsample includes all AGN with radial velocity in the range  
[1500$\div$3000] km$s^{-1}$: 48 LINERs (including transition objects), 
15 Seyfert galaxies and 52 HII galaxies. 
  
For each AGN we have automatically identified neighbours in the UZC catalog 
(Falco et al. 1999) which is complete for galaxies brighter than $m_{zw}$=15.5
The local density of each AGN is estimated by counting all neighbours within a cylinder 0.2 $h^{-1}$ Mpc in projected radius and $\pm$1000 km/s in depth.
The large scale density is estimated by counting all neighbours within a cylinder 1 $h^{-1}$ Mpc in projected radius and $\pm$1000 km$s^{-1}$ in depth.
The distance to the nearest among these neighbours defines the nearest neighbour distance. 
%
%                                                Figure with two panels
%----------------------------------------------------------- S_vib 
   \begin{figure}
   \centering
\includegraphics[clip=true,angle=0,width=0.55\hsize]{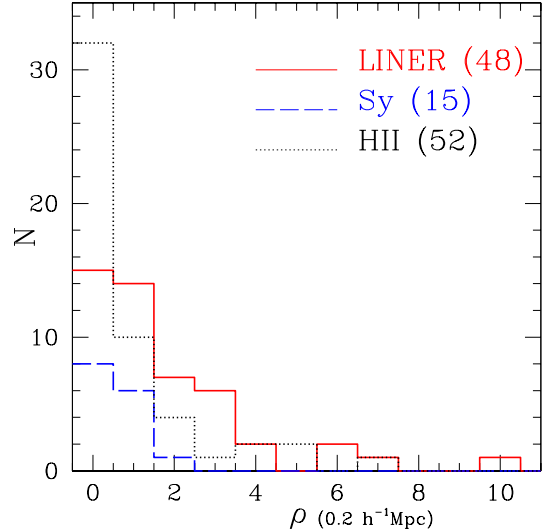}
     \caption{Local density distributions (neighbours within 200 $h^{-1}$kpc) of LINERs, Seyferts and HII galaxies. LINERs exhibit significantly larger local density than Seyfert and HII galaxies. }  
        \label{fit1}
    \end{figure}
   \begin{figure}
   \centering
%\resizebox{\hsize}{!}{\includegraphics[clip=true]{zitelli_fig1.eps}
\includegraphics[clip=true,angle=0,width=0.55\hsize]{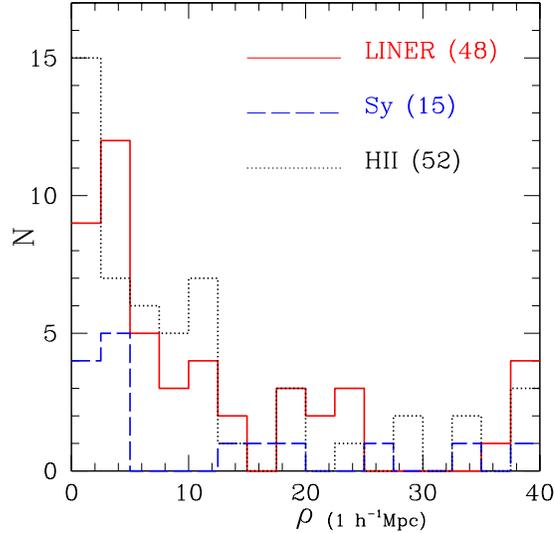}
     \caption{Large scale density distributions of LINERs, Seyferts and HII galaxies. The hyphothesis that LINERs, Seyfert and HII galaxies are drawn from the same population cannot be rejected at the 1$\sigma$ significance level          }  
        \label{fit2}
    \end{figure}
%______________________________________________________________
\section {Neighbour density}
In fig.\ref{fit1} we show the distribution of AGNs as a function of the 
local neighbour density (neighbours within 0.2 $h^{-1}$Mpc) 
for Seyferts, Liners and HII galaxies respectively. 
Distributions show some difference by eye.  
The U-test indicates,  at a significance level 3$\sigma$, 
that the local density distributions 
of LINERs and Seyferts are drawn from different 
parent distributions.  
Significant differences (3$\sigma$) are also found when comparing 
(U-test) the local density of LINERs and HII galaxies,  
whereas Seyferts and HII galaxies do not appear to differ significantly. 

In fig.\ref{fit2} we show the distribution of AGNs as a function of the 
large scale neighbour density (neighbours within 1 $h^{-1}$Mpc).  
The large scale density distributions of LINERs, Seyferts and HII galaxies 
appear similar. The U- test indicates that differences between 
population are not significant. 
\section{Nearest Neighbour Distance}
%-------------------------------------------------------------
   \begin{figure}
   \centering
\includegraphics[angle=0,width=0.55\hsize]{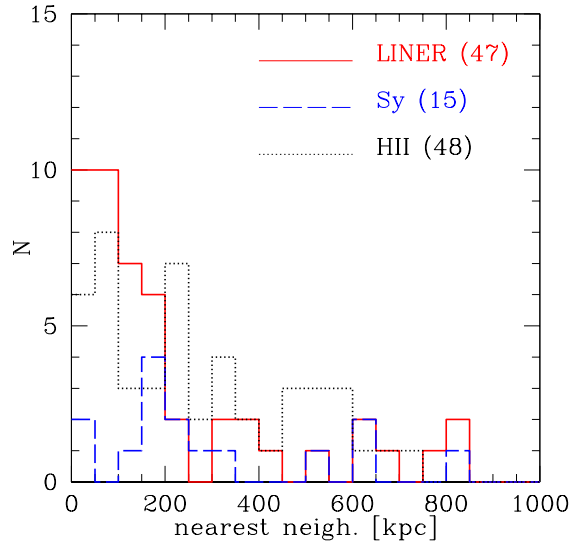}
     \caption{Distribution of AGNs as a function of the distance to the nearest neighbour. LINERs exhibit nearest neighbours that are closer  than 
those of Seyfert and HII galaxies. }  
        \label{fit3}
    \end{figure}
The distribution of LINERs, Seyferts and HII galaxies as a function of the 
distance to the nearest neighbour is shown in fig \ref{fit3}. 
The plot suggests that Liners exhibit nearest neighbours that are closer than 
those of Seyfert and HII galaxies. The KS indicates, at a 2$\sigma$ 
significance level, that distributions of LINERs and of Seyfert or HII 
galaxies  are drawn from different parent populations.  
Conversely, we find no evidence that Seyferts and HII galaxies 
are drawn from different population.  
\section{Elliptical and spiral hosts}
There is a tendency for LINERs to be found in richer and more massive 
Compact Groups than Seyfert galaxies (Kelm et al. 2004). 
LINERs are further found in galaxies of earlier Hubble type than Seyferts and 
HII galaxies, suggesting that the excess of neighbours around LINERs 
 might be due to a morphology density effect (Schmitt 2001).  
To test whether this effect is responsible for the excess of close 
companions that we find around LINERs we have excluded from our sample all
AGNs located in early-type hosts. This reduces the size of samples  
to 28 LINERs, 11 Seyfert galaxies and 49 HII galaxies. 

Indeed, when comparing the nearest neighbour distribution 
of late-type galaxy samples we find that differences between 
LINERs and other AGNs turn out to be non significant. 
And differences in local density distributions between LINERs and Seyfert or 
HII galaxies  become less significant. 

We therefore confirm that the excess of neigbours around LINERs 
is most likely due to a morphology density effect. 
\bibliographystyle{aa}

\end{document}